\documentclass[aps,prep,epsfile,nofootinbib]{revtex4}

\usepackage{graphics}
\usepackage{slashed}
\usepackage{amssymb}
\usepackage{amsmath}
\usepackage{graphicx}
\usepackage{epsfig}

\begin{document}

\title{Spectral dimensionality of spacetime around a radiating Schwarzschild black-hole}
\author{$^{1,2}$ Mauricio Bellini\footnote{mbellini@mdp.edu.ar}, $^{1,2}$ Juan Ignacio Musmarra\footnote{jmusmarra@mdp.edu.ar}, $^{1,2}$ Pablo Alejandro S\'anchez\footnote{pabsan@mdp.edu.ar} and $^{1}$ Alan S. Morales\footnote{alanmorales@mdp.edu.ar}}
\address{
$^1$ Instituto de Investigaciones F\'{\i}sicas de Mar del Plata - IFIMAR, \\
Consejo Nacional de Investigaciones Cient\'ificas y T\'ecnicas
- CONICET, Mar del Plata, Argentina.\\
$^2$ Departamento de F\'{\i}sica, Facultad de Ciencias Exactas y
Naturales, Universidad Nacional de Mar del Plata, Funes 3350, C.P. 7600, Mar del Plata, Argentina.}
\begin{abstract}
In this work we study the spectral dimensionality of spacetime around a radiating Schwarzschild black hole using a recently introduced formalism of quantum gravity, where the alterations of the gravitational field produced by the radiation are represented on an extended manifold, and describe a non-commutative and non-linear algebra. The ration between classical and quantum perturbations of spacetime can be measured by the parameter $z \geq 0$. When $z=(1+\sqrt{3})/2\simeq 1.3660$, a relativistic observer approaching the Schwarzschild horizon perceives a spectral dimension $N(z)=4\left[\theta(z)-1\right]\simeq 2.8849$. Under these conditions, all studied Schwarzschild black holes with masses ranging from the Planck mass to $10^{46}$ times the Planck mass, present the same stability configuration which suggests the existence of an universal property of these objects under those particular conditions. The difference from the spectral dimension previously obtained at cosmological scales leads to the conclusion that the dimensionality of spacetime is scale-dependent. Another important result presented here, is the fundamental alteration of the effective gravitational potential near the horizon due to Hawking radiation. This quantum phenomenon prevents the potential from diverging to negative infinity as the observable approaches the Schwarzschild horizon.
\end{abstract}
\maketitle

\section{Introduction and motivation}

Employing a geometrodynamical description, quantum gravity delves into the quantum characteristics of spacetime, considering geometry as a dynamic quantum entity \cite{Carlip2001,Addazi2021}. This approach aims to understand spacetime behavior at the quantum level, potentially leading to discrete quantum geometries and a fundamentally dynamic description of the universe at its smallest scales \cite{Hossenfelder2013}. In this context, spacetime geometry can exist in a superposition of states, analogous to quantum particles \cite{Belenchia2018}. Consequently, at extremely small scales, spacetime itself may exhibit quantum fluctuations \cite{Smolin1986}. This contrasts with classical general relativity, where spacetime is modeled as a smooth and continuous manifold.

There are several theoretical frameworks which currently attempt to describe gravity at the quantum level. Some interesting examples include M-theory \cite{BB1,BB2} and various approaches within loop quantum gravity, such as spin foam models \cite{BB3,BB4,BB5}. A common feature among these theories is their reliance on geometric structures. Quantum geometrodynamics \cite{BB6} and quantum geometry \cite{BB4,BB5} exemplify some of the geometric approaches used in this context. A more recent development addressing this challenge is the formulation of Relativistic Quantum Geometry \cite{M1,M2}, which constructs a gauge-invariant relativistic quantum geometric framework based on a Weyl-like manifold. This formalism has been employed in recent studies of backreaction effects at both astrophysical \cite{Musmarra:2023dfl,Bellini:2024meo} and cosmological \cite{Bellini:2023ddy} scales, yielding significant results that may offer novel insights into the pursuit of a quantum theory of gravity. The goal of this approach is to understand the spacetime behavior at the quantum level, potentially leading to the emergence of discrete quantum geometries and a fundamentally dynamic description of the universe at most fundamental scales \cite{Hossenfelder2013}. In this context, spacetime geometry can exist in a superposition of states, similar to quantum particles \cite{Belenchia2018}. As a result, spacetime itself may exhibit quantum fluctuations at extremely small scales \cite{Smolin1986}, which stands in contrast to the smooth and continuous manifold model of classical general relativity. These questions are especially relevant in the context of primordial black holes \cite{Dong:2015yjs}, whose complete evaporation could leave observable imprints or even remnants, depending on the underlying theory. Moreover, several of these proposals consider or predict that the dimensionality of spacetime might not be a fixed quantity, but rather could flow toward values close to $2$ at very short distances \cite{Carlip:2017eud,Carlip:2019onx,Saueressig:2021pzy}, or even take on complex values \cite{Calcagni:2017via}, potentially leading to observationally relevant consequences \cite{Calcagni:2019ngc}.

If we consider a radiating Schwarzschild black hole (SBH), the emitted radiation induces geometric alterations in the spacetime surrounding the SBH. In the context of RQG, these alterations are regarded as quantum in nature and are understood to modify the geometry of the Riemannian background, which is initially described by the metric tensor of an unperturbed SBH. Moreover, this process gives rise to a geometric flow from the SBH to its surroundings, which can be captured by variations in the Einstein–Hilbert action. The information associated with this flow is encoded in the cosmological parameter $\lambda$, defined via a proportionality between perturbations of the metric tensor and those of the Ricci tensor. Its exact functional form can be determined by considering an invariant associated with the relativistic velocity $\bar{U}^{\alpha}$, based on the ratio $z = b/\eta$, where $b$ and $\eta$ are free parameters that characterize the deviation of the connections on the extended manifold from the Levi-Civita connections. Within this framework, the inclusion of perturbations also leads to a change in the effective number of spacetime dimensions, denoted by $N(z)$. As we shall show, this behavior is consistent with various approaches to quantum gravity \cite{Carlip2011,Calcagni2017}, particularly those in which the spectral dimension of the Schwarzschild spacetime decreases continuously at shorter distance scales.

In this work, we shall study the geometrical dynamics of the environment in a SBH, produced by the flow of geometric quantum fluctuations of spacetime in the vicinity of the Schwarzschild horizon, by using the formalism introduced in two previous works \cite{p1,p2}. The structure of the paper is as follows. In Sect. II, we provide a brief overview of the Relativistic Quantum Geometry formalism and explicitly present the computation of the metric tensor perturbations and the cosmological parameter in terms of the relativistic velocities $\bar{U}^{\alpha}$ and the ratio $z$. In Sect. III, we compute the quantum-induced geometric fluctuations in the vicinity of a SBH by solving the geodesic equation for the case $z = (1 + \sqrt{3})/2$ and $\bar{U}^2 = \bar{U}^3 = 0$. We then analyze the effects of this contribution through plots of the cosmological parameter and the effective gravitational potential for different mass values, as well as the spectral dimensionality and the quantum corrections to the metric tensor. In Sect. IV, we offer some final remarks concerning the results obtained and possible future developments.

\section{Relativistic Quantum Geometry}

\begin{figure}
    \centering
    \includegraphics[scale=0.5]{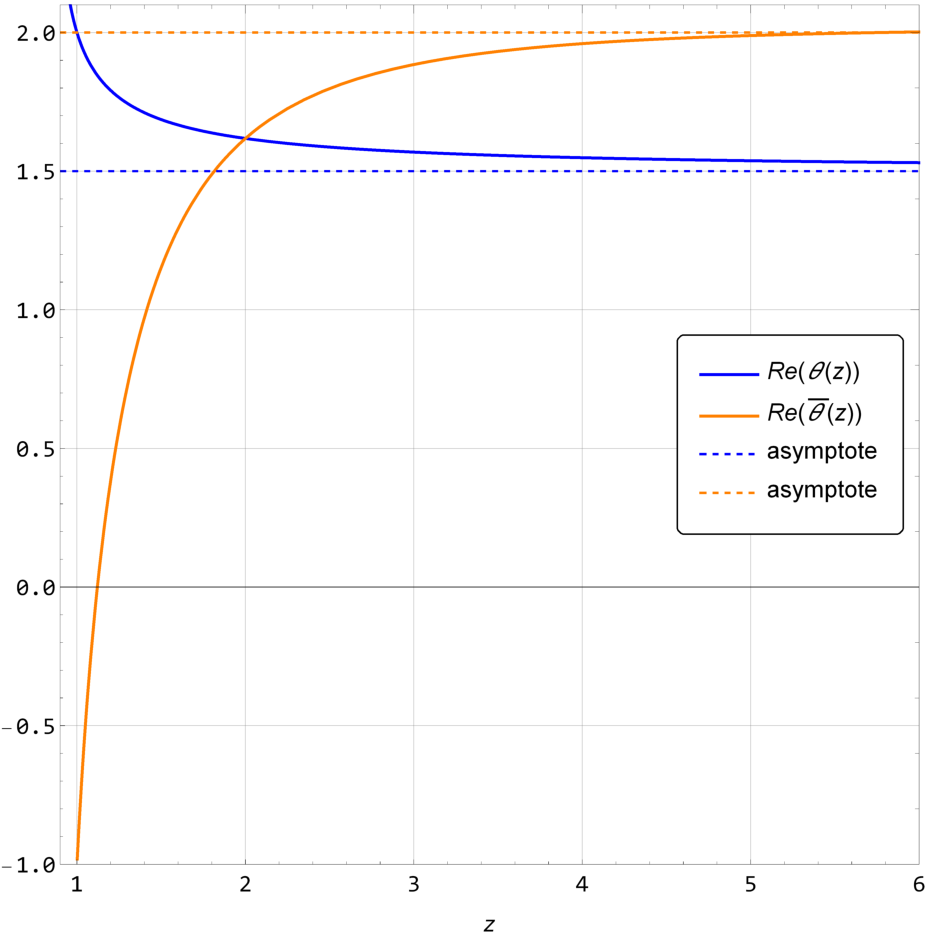}
    \caption{Real contributions of $\theta(z)$ (blue line) and $\bar\theta(z)$ (orange line) for $z> 0.9$.}
    \label{f1gi147}
\end{figure}

We shall consider an Einstein-Hilbert (EH) action which is considered from its quantum nature. The integrand in the action can be considered as the expectation value on a Riemann manifold of quantum magnitudes, such that the Lagrangian density $\hat{\cal L}_m$
\begin{equation}\label{action}
{\cal I} = \int\,d^4x\,\sqrt{-g}\,\left\{\frac{R}{2\kappa}+\left<B\right|\hat{\cal L}_m\left|B\right>\right\},
\end{equation}
where $\kappa= 8\pi\,G/c^4$. The background expectation value on the Riemann manifold of $\hat{\cal L}_m$: $\left<B\right|\hat{\cal L}_m\left|B\right>$, is calculated with a quantum state $\left|B\right>$, by using the Heisenberg representation, where quantum operators evolve in spacetime, while the quantum states are squeezed in space and time. If $R_{\alpha\beta}$ is the classical Ricci tensor and $g_{\alpha\beta}$ are the components of the metric tensor with determinant $g$ and signature $(+,-,-,-)$, then the scalar curvature will be defined by $R=g^{\alpha\beta}\,R_{\alpha\beta}$. This means that the covariant derivatives on this manifold will be defined with respect to the Levi-Civita connections \begin{tiny}$\left\{ \begin{array}{cc}  \alpha \, \\ \beta \, \nu  \end{array} \right\}$\end{tiny}. The Einstein equations on the background metric are obtained after varying the EH action (\ref{action})
\begin{equation}\label{delta0}
\delta {\cal I} = \int d^4 x \sqrt{-g} \left[{\delta g}^{\alpha\beta}\,G_{\alpha\beta}+  \left<B\right|\kappa\,\delta \hat{g}^{\alpha\beta}   \hat{T}_{\alpha\beta}+ \,g^{\alpha\beta} \hat{\delta R}_{\alpha\beta}\left|B\right> \right]=0.
\end{equation}
The last terms in (\ref{delta0}) will be considered of quantum nature:
\begin{equation}\label{RR}
\delta\Theta\equiv\left<B\right|\hat{\delta\Theta}\left|B\right>=\left<B\right|g^{\alpha\beta} \hat{\delta R}_{\alpha\beta}\left|B\right>,
\end{equation}
with $\hat{\delta g}_{\alpha\beta}$ as geometrical sources of $\hat{\delta R}_{\alpha\beta}$ \cite{mabe}
\begin{equation}\label{fff}
\hat{\delta R}_{\alpha\beta}= \lambda\left(x^{\mu}\right)\,\hat{\delta g}_{\alpha\beta}.
\end{equation}
Furthermore, $\hat{\delta\Theta}=\left(\hat{\delta W}^{\alpha}\right)_{\|\alpha}$ is the quantum flux of the $4$-vector field
\begin{equation}\label{W}
\hat{\delta W}^{\alpha}=b\,\left(\hat{\delta\Gamma}^{\epsilon}_{\beta\epsilon}\, {g}^{\beta\alpha}-\hat{\delta \Gamma}^{\alpha}_{\beta\gamma}\, {g}^{\beta\gamma}\right),
\end{equation}
whose expectation value given by (\ref{RR}), alters the classical relativistic dynamics on the Riemann manifold and this modification is reflected through $\lambda\left(x\right)$, knowing as the cosmological parameter in a cosmological context \cite{p2}. The classical variation of the connections is given the expectation value on the Riemann manifold: $\delta\Gamma^{\mu}_{\beta\epsilon}= \left< B\right|\hat{\delta\Gamma}^{\mu}_{\beta\epsilon}\left|B\right>=b\,\bar{U}^{\mu}\,g_{\beta\epsilon}$ and $\hat{\delta g}_{\alpha\beta}=\hat{g}_{\alpha\beta\|\mu}\,\bar{U}^{\mu}$, such that $\bar{U}^{\mu}=\frac{dx^{\mu}}{dS}$ are the relativistic velocity components with the geometric fluctuations included, and are solutions of the differential equations \cite{p1}
\begin{equation}\label{geo2}
\frac{d \bar{U}^{\alpha}}{d S} + \Gamma^{\alpha}_{\mu\beta} \,\bar{U}^{\mu}\,\bar{U}^{\beta} = \left[ \eta-b\right] \,\bar{U}^{\alpha}\,\left[\theta(z)-1\right],
\end{equation}
when we have used the fact that $\bar{U}_{\mu}\bar{U}^{\mu}=g_{\nu\mu}\,\bar{U}^{\nu}\bar{U}^{\mu}=\left[\theta(z)-1\right]$,
with $\theta(z)$ given by \cite{p1}
\begin{equation}\label{oro}
\theta(z)= \frac{\left[3z-4+\sqrt{9\,z^2-8\,z}\right]}{4\,(z-1)},
\end{equation}
for $b=z\,\eta$. The covariant derivative of the metric tensor on the extended manifold\footnote{The covariant derivative of a $(n)$-times contravariant and $(m)$-times covariant mixture tensor field $\hat{\Upsilon}^{\alpha_1...\alpha_n}_{\beta_1 ... \beta_m}$, on the extended manifold, is given by the $(n+m+1)$-range quantum tensor:
\begin{eqnarray}
&& \hat{\Upsilon}^{\alpha_1...\alpha_n}_{\beta_1 ... \beta_m \|\mu} = \hat{\nabla}_{\mu} \hat{\Upsilon}^{\alpha_1...\alpha_n}_{\beta_1 ... \beta_m} +
\sum_{i=1}^{n} \hat{\delta\Gamma}^{\alpha_i}_{\nu\mu}\,\hat{\Upsilon}^{\alpha_1..\alpha_{i-1}\nu\alpha_{i+1} ..\alpha_n}_{\beta_1 ... \beta_m}
-\sum_{i=1}^{m} \hat{\delta\Gamma}^{\nu}_{\mu \beta_i}\,\hat{\Upsilon}^{\alpha_1...\alpha_n}_{\beta_1 ..\beta_{i-1}\nu\beta_{i+1}. \beta_m}  \nonumber \\
&& -\eta \sum_{i=1}^{n-1} \left(\hat{\Upsilon}^{\alpha_1..\alpha_i\alpha_{i+1}..\alpha_n}_{\beta_1 ... \beta_m}\,\hat{\Omega}_{\mu}
+\hat{\Omega}_{\mu}\,\hat{\Upsilon}^{\alpha_1..\alpha_{i+1}\alpha_i..\alpha_n}_{\beta_1 ... \beta_m}\right) + \eta \sum_{i=1}^{m-1}\left(\hat{\Upsilon}^{\alpha_1...\alpha_n}_{\beta_1 ..\beta_i\beta_{i+1}.. \beta_m}\, \hat{\Omega}_{\mu}+
\hat{\Omega}_{\mu}\,\hat{\Upsilon}^{\alpha_1...\alpha_n}_{\beta_1 ..\beta_{i+1}\beta_i.. \beta_m}\right), \label{uau}
\end{eqnarray}
where the terms in the last row of (\ref{uau}) describe the interaction of the quantum tensor $\hat{\Upsilon}^{\alpha_1...\alpha_n}_{\beta_1 ... \beta_m}$ with the extended manifold. Furthermore, $\eta$ is a parameter related to the quantum effects and $b=z\eta$ is related to the classical ones \cite{p2}. Additionally, we define the quantum variation of the quantum tensor field $\hat{\Upsilon}^{\alpha_1...\alpha_n}_{\beta_1 ... \beta_m }$:
\begin{displaymath}
\hat{\delta\Upsilon}^{\alpha_1...\alpha_n}_{\beta_1 ... \beta_m}=\hat{\Upsilon}^{\alpha_1...\alpha_n}_{\beta_1 ... \beta_m \|\mu}\,\bar{U}^{\mu}.
\end{displaymath}}
is
\begin{equation}
\hat{g}_{\alpha\beta\|\mu} = \eta\left[2\,g_{\alpha\beta}\,\hat{\Omega}_{\mu} - z \left(\hat{\Omega}_{\beta}\,g_{\alpha\mu}+\hat{\Omega}_{\alpha}\,g_{\beta\mu}\right)\right],
\end{equation}
where we have considered $\nabla_{\mu} g_{\alpha\beta}=0$, in agreement with the expected for the metric tensor on a Riemann manifold. In this work, as in previous ones \cite{p1,p2}, we shall consider the case where
$\hat{\delta\Gamma}^{\epsilon}_{\beta\epsilon}=b\,\hat{\Omega}^{\mu}\,g_{\beta\epsilon}$, such that $\left<B\right|\hat{\Omega}^{\mu}\left|B\right>=\bar{U}^{\mu}$, $\left<B\right|\hat{\Omega}_{\mu}\left|B\right>=\bar{U}_{\mu}$ and $\left<B\right|\hat{\Omega}_{\mu}\hat{\Omega}^{\mu}\left|B\right>=\left[\bar{\theta}(z)-1\right]$, with \cite{p1}
\begin{equation}\label{tteta}
\bar{\theta}(z)= \frac{4z^3-3z^2-10z+8+\sqrt{9z^2-8z}(2-z)}{z(z-1)\left[\sqrt{9z^2-8z}-z\right]}.
\end{equation}
In the Fig. (\ref{f1gi147}) we have plotted the real contributions of $\theta(z)$ (blue line) and $\bar{\theta}(z)$ (red line), for $z> 0.9$. Notice that $\lim_{z\rightarrow \infty} \theta(z) \rightarrow 3/2$ and $\lim_{z \rightarrow \infty} \bar{\theta}(z) \rightarrow 2$.

\subsection{Quantum geometric description of Gravity}

The quantum geometric field $\hat{\Omega}$, that describes the quantum gravitational fluctuations with respect to the background, can be represented by means of a Fourier expansion
\begin{equation}\label{fu}
\hat{\Omega}(\tau,\vec{r}) = \frac{1}{(2\pi)^{3/2}}\,\int \,d^3 k\,\left[ \hat{\Omega}_k(\tau,\vec{r})+\hat{\Omega}^{\dagger}_k(\tau,\vec{r})\right].
\end{equation}
The modes of the expansion are $\hat{\Omega}_k(\tau,\vec{r})=\hat{A}_k\,{\sigma}(k,\tau,\vec{r})$ and $\hat{\Omega}^{\dagger}_k(\tau,\vec{r})=\hat{A}^{\dagger}_k\,{\sigma}^*(k,\tau,\vec{r})$.  The creation and destruction operators: $\hat{A}^{\dagger}_k$ and $\hat{A}_k$, comply with the algebra
\begin{equation}\label{m55}
\left<B\left|\left[\hat{A}_k(\tau,\vec{r}),\hat{A}^{\dagger}_{k'}(\tau,\vec{r})\right]\right|B\right>
=i\,\delta^{(3)}\left(\vec{k}-\vec{k'}\right),\quad
\left<B\left|\left[\hat{\Omega}^{\dagger}_k(\tau,\vec{r}),\hat{\Omega}^{\dagger}_{k'}(\tau,\vec{r})\right]\right|B\right>=
\left<B\left|\left[\hat{\Omega}_k(\tau,\vec{r}),\hat{\Omega}_{k'}(\tau,\vec{r})\right]\right|B\right>=0.
\end{equation}
Here, $\hat{\Omega}$ is a scalar field and we are denoting $\hat{\partial}_{\mu}\hat{\Omega}\equiv \hat{\Omega}_{\mu}$, such that $\hat{\partial}_{\mu}$ must be understood as a differential operator applied to the operator $\hat{\Omega}$ that complies with the nonlinear and noncommutative algebra \cite{p1}
\begin{equation}\label{m56}
\left[\frac{\hat{\partial}}{\partial{x}^{\alpha}},\hat{\Omega}_{k'}^\dagger\right]=-i\,\bar{U}_{\alpha}\,
\left[\hat{\Omega}_{k},\hat{\Omega}_{k'}^\dagger\right].
\end{equation}
The right-hand term in (\ref{m56}) alters the standard geodesic equations when $b \neq \eta$.

The classical variations of the metric tensor components are given by
\begin{equation}\label{delg}
\delta g_{\alpha\beta} = b^{-1}\,\left<B\right|\hat{\delta g}_{\alpha\beta}\left|B\right>,
\end{equation}
and the total metric tensor components with quantum gravitational fluctuations included are given by
\begin{equation}\label{gbar}
\bar{g}_{\alpha\beta}=  g_{\alpha\beta} + \delta g_{\alpha\beta} = g_{\alpha\beta} \left[1+\frac{2}{z} \left[\theta(z)-1\right]-\bar{U}_{\alpha}\bar{U}_{\beta}\right].
\end{equation}
In this way, since $\bar{g}^{\alpha\beta}=  g^{\alpha\beta} + \delta g^{\alpha\beta}$, can be obtained that the effective number of dimensions of spacetime with quantum gravitational fluctuations included, $N(z)$: \cite{p1}
\begin{equation}
N(z) \equiv \bar{g}_{\alpha\beta}\bar{g}^{\alpha\beta} = 4 \left[\theta(z)-1\right],
\end{equation}
and $\bar{g}_{\alpha\beta}\,\bar{U}^{\alpha}\bar{U}^{\beta}=1$. From (\ref{RR}) and (\ref{fff}) we obtain the expression for $\lambda(x^{\mu})$: \cite{p1,p2})
\begin{equation}\label{lam}
\lambda(x^{\nu}) = \frac{3z\,\eta}{2} \left[\frac{\eta\,(2z+1)\,\left[\bar{\theta}(z)-1\right] - g^{\alpha\beta} \Gamma^{\nu}_{\alpha\beta} \bar{U}_{\nu}}{ (z-1) \left[\theta(z)-1\right]}\right].
\end{equation}
Note that the first term of $\lambda(x^{\nu})$ is purely due to geometric quantum contributions, and the second term has purely classical relativistic contributions. In this work we are interested to study the value of $\lambda(x^{\nu})$ around the Schwarzschild radius in a SBH. To achieve this, we must solve the geodesic equations (\ref{geo2}), which incorporate the effects of quantum fluctuations in the gravitational field via the source term on the right-hand side. Due to the manifest isotropy of a SBH, we shall focus on solutions to equations (\ref{geo2}) where
$\bar{U}^2=\bar{U}^3=0$.

\subsection{Background Einstein equations with geometric quantum gravitational corrections included}

When the geometric quantum gravitational corrections are included the Einstein equations must be modified as \cite{p2}
\begin{equation}
G_{\alpha\beta} =-\kappa\,\left[T_{\alpha\beta}- g_{\alpha\beta}\,\frac{\lambda(x^{\mu})}{\kappa}\right],
\end{equation}
where ${\delta g}^{\alpha\beta} \bar{T}_{\alpha\beta}= \left<B\right|\hat{\delta g}^{\alpha\beta}  \left[ \hat{T}_{\alpha\beta} -  \frac{{\lambda}(x^{\nu})}{\kappa}\,{ g}_{\alpha\beta}\right]\left|B\right>$ in Eq. (\ref{delta0}), is the expectation value on the Riemann manifold of the redefined stress tensor, with Eq. (\ref{fff}) and $g_{\alpha\beta} \,\delta \hat{g}^{\alpha\beta} + g^{\alpha\beta}\,\delta \hat{g}_{\alpha\beta}=0$:
\begin{equation}
\bar{T}_{\alpha\beta}=T_{\alpha\beta}- g_{\alpha\beta}\,\frac{\lambda(x^{\mu})}{\kappa},
\end{equation}
and ${{T}}_{\alpha\beta}= 2 \left<B\right|\frac{\delta {\hat{\cal L}_m}}{\delta \hat{g}^{\alpha\beta}}\left|B\right>  - g_{\alpha\beta} \left<B\right|{\hat{\cal L}_m}\left|B\right>$, for the Lagrangian density $\hat{\cal L}_m$. The field equations on the Riemann manifold, describe the dynamics of the system, and are
\begin{equation}
\nabla_{\alpha} G^{\alpha\beta}=\nabla_{\alpha} \bar{T}^{\alpha\beta}=0.
\end{equation}
Furthermore, the Einstein tensor $G_{\alpha\beta}$ is given by
\begin{equation}
G_{\alpha\beta}= R_{\alpha\beta} - \frac{1}{2}\,g_{\alpha\beta}\,R,
\end{equation}
such that $R_{\alpha\beta}$ is the Ricci tensor defined in the standard manner in terms of the Levi-Civita connections $\left\{ \begin{array}{cc}  \nu \, \\ \alpha \, \mu  \end{array} \right\}$
\begin{displaymath}
R_{\alpha\beta} \equiv R^{\nu}_{\alpha\beta\nu}= \left\{ \begin{array}{cc}  \nu \, \\ \alpha \, \nu  \end{array} \right\}_{,\beta}-\left\{ \begin{array}{cc}  \nu \, \\ \alpha \, \beta  \end{array} \right\}_{,\nu} + \left\{ \begin{array}{cc}  \nu \, \\ \beta \, \mu  \end{array} \right\} \left\{ \begin{array}{cc}  \mu \, \\ \nu \, \alpha  \end{array} \right\} -
\left\{ \begin{array}{cc}  \mu \, \\ \mu \, \nu  \end{array} \right\} \left\{ \begin{array}{cc}  \nu \, \\ \alpha \, \beta  \end{array} \right\} ,
\end{displaymath}
and $R=g^{\alpha\beta}\,R_{\alpha\beta}$ is the scalar of curvature.

\section{Quantum geometric fluctuations of spacetime around the SBH}

We consider the Schwarzschild line element with coordinates $x^0=\tau$, $x^2=r$, $x^2=\vartheta$ and $x^3=\varphi$:
\begin{equation}\label{m1}
dS^2 = f(r)\,d\tau^2 -\left[\frac{dr^2}{f(r)}  + r^2\,\left(d\vartheta^2 + \sin^2(\vartheta)\,d\varphi^2\right)\right],
\end{equation}
with $\tau=c\,t$  $f(r)=1-2\,M/r$ and $M=G\,m/c^2$. Here, $t$ is the time, $G$ is the gravitational constant, $m$ is the mass of the SBH and $c$ is the light velocity.

Since the exterior of a SBH describes a vacuum, all components of the Einstein tensor are zero: $G_{\alpha\beta}=0$. Consequently, the modified stress tensor components will also be zero: $\bar{T}_{\alpha\beta}=0$. For a perfect fluid incorporating the expectation value of quantum gravitational fluctuations, this tensor is given by
\begin{equation}\label{stress}
\bar{T}^{\alpha}_{\hskip .2cm\beta} = \left(P+\rho\,c^2\right)\,\bar{U}^{\alpha}\bar{U}_{\beta} - \delta^{\alpha}_{\hskip .2cm\beta} \,P - \delta^{\alpha}_{\hskip .2cm\beta}\, \frac{\lambda(x^{\nu})}{\kappa}=0,
\end{equation}
where $P$ is the pressure and $\rho$ is the mass density. When the background geometry is altered by a classical and quantum fluctuations of spacetime, the relativistic velocities comply with \cite{p1}:
\begin{equation}
\bar{U}^{\alpha}\bar{U}_{\beta}=  \delta^{\alpha}_{\hskip .2cm\beta} \left[\theta(z) - 1\right].
\end{equation}
Therefore, if we consider an equation of state given by $\Omega=P/(\rho\,c^2)$, we obtain from (\ref{stress}) that
\begin{equation}
\Omega = \frac{1}{\left[\theta(z)-2\right]} \left\{\frac{\lambda(x^{\nu})}{\kappa\,\rho\,c^2}-\left[\theta(z)-1\right] \right\},
\end{equation}
which provides the equation of state in the environment of a SBH when quantum geometric fluctuations of spacetime are taken into account.

\subsection{Calculation of $\lambda(x^{\nu})$ around the SBH}

The Levi-Civita connections are given by
\begin{equation}
\left\{ \begin{array}{cc}  \alpha \, \\ \mu \, \nu  \end{array} \right\} = \frac{1}{2}\,g^{\alpha\sigma} \left[\partial_{\nu}\, g_{\sigma\mu}+\partial_{\mu}\, g_{\sigma\nu}  - \partial_{\sigma}\, g_{\mu\nu}\right].
\end{equation}
The relevant ones, for the metric (\ref{m1}), are
\begin{eqnarray}
&& \left\{ \begin{array}{cc}  0 \, \\ 1 \, 0  \end{array} \right\} = \frac{M}{r(r-2M)}, \qquad
\left\{ \begin{array}{cc}  1 \, \\ 0 \, 0  \end{array} \right\}= \frac{M(r-2M)}{r^3},
\qquad
\left\{ \begin{array}{cc}  1 \, \\ 1 \, 1  \end{array} \right\}= -\frac{M}{r(r-2M)}, \\
&& \left\{ \begin{array}{cc}  1 \, \\ 2 \, 2  \end{array} \right\}=-(r-2M), \qquad \left\{ \begin{array}{cc}  1 \, \\ 3 \, 3  \end{array} \right\}=
-(r-2M)\,\sin^2(\vartheta), \qquad
\left\{ \begin{array}{cc}  2 \, \\ 1 \, 2  \end{array} \right\}=\frac{1}{r}, \\
&& \left\{ \begin{array}{cc}  2 \, \\ 3 \, 3  \end{array} \right\}=-\sin(\vartheta)\,\cos(\vartheta), \qquad
\left\{ \begin{array}{cc}  3 \, \\ 1 \, 3  \end{array} \right\}=\frac{1}{r}, \qquad
\left\{ \begin{array}{cc}  3 \, \\ 2 \, 3  \end{array} \right\}=\frac{\cos(\vartheta)}{\sin(\vartheta)}.
\end{eqnarray}
The calculation in the second term of (\ref{lam}), is given by
\begin{equation}
 g^{\alpha\beta} \Gamma^{\nu}_{\alpha\beta} \bar{U}_{\nu} = - \bar{U}_1(S)/r.
\end{equation}
Furthermore, we shall assume that $\eta=1/\left(2M\right)$. In the following items we shall solve the geodesic equations (\ref{geo2}), for
the relativistic components $\bar{U}^0(S)$ and $\bar{U}^1(S)$. To do this, we must take into account the normalization condition for the velocity
components:
\begin{equation}\label{norma}
g_{00}\,\left[\bar{U}^0\right]^2+g_{11}\,\left[\bar{U}^1\right]^2= \theta(z)-1,
\end{equation}
for $\theta(z)\geq 1$. In the Fig. (\ref{f1gi147}), can be viewed that $\theta(z) >1$ for $z>8/9$.

\subsubsection{Solution for $\bar{U}^0(S)$}

To solve the differential equation for $\bar{U}^0(S)$
\begin{equation}
\frac{d\bar{U}^0}{dS} + \left\{ \begin{array}{cc}  0 \, \\ \alpha \, \beta  \end{array} \right\}\,\bar{U}^{\alpha}\bar{U}^{\beta}+
\eta\,\left[\theta(z)-1\right]\,(z-1)\,\bar{U}^0=0,
\end{equation}
we can define the following auxiliary component: $\tilde{U}^0 =g_{00}\,\bar{U}^0$, and after some algebra, we obtain
\begin{equation}
\frac{d}{dS} \tilde{U}^0 + \eta\,(z-1)\,\left[\theta(z)-1\right]\, \tilde{U}^0=0,
\end{equation}
which has the following solution
\begin{equation}
 \tilde{U}^0=  C\,e^{- \eta\,(z-1)\,\left[\theta(z)-1\right]\,S},
\end{equation}
where $C\equiv \tilde{U}^0(S_0)=1$, for $S_0=0$. Therefore, the solution for $\bar{U}^0(S)$ will be given by
\begin{equation}\label{u0}
\bar{U}^0(S)= \left[\frac{r(S)}{r(S)-2M}\right]\,e^{- \eta\,(z-1)\,\left[\theta(z)-1\right]\,S},
\end{equation}
which, for $S\geq 0$, $\eta>0$, $r(S)\geq 2M$ and $z>1$, decreases as the system evolves.

\subsubsection{Solution for $\bar{U}^1(S)$}

The geodesic equation for $\bar{U}^1(S)$
\begin{equation}
\frac{d\bar{U}^1}{dS} + \left\{ \begin{array}{cc}  1 \, \\ \alpha \, \beta  \end{array} \right\}\,\bar{U}^{\alpha}\bar{U}^{\beta}+
\eta\,\left[\theta(z)-1\right]\,(z-1)\,\bar{U}^1=0,
\end{equation}
has the explicit form
\begin{eqnarray}
&& \frac{d\bar{U}^1}{dS} + \left[\frac{2M}{r(r-2M)}\right] \,\bar{U}^0\,\bar{U}^1-\left[\frac{M}{r(r-2M)}\right] \,\left[\bar{U}^1\right]^2 \nonumber \\
&& + \left[\frac{2M\,(r-2M)}{r^3} \right] \,\left[\bar{U}^0\right] ^2 +\eta\,\left[\theta(z)-1\right]\,(z-1)\,\bar{U}^1=0.
\end{eqnarray}
From the equation (\ref{norma}), we obtain that
\begin{equation}\label{c1}
\frac{d\bar{U}^1}{dS} +\left[ \frac{2M}{r(r-2M)} \right]\,\bar{U}^0\,\bar{U}^1 +
\eta\,\left[\theta(z)-1\right]\,(z-1)\,\bar{U}^1+\left[\frac{M}{r^2}\right]\,\left[\theta(z)-1\right]=0.
\end{equation}
We can derive (\ref{norma}) with respect to $S$, and using the fact that $r\equiv r(S)$ and the expression (\ref{u0}), we obtain that
\begin{equation}\label{c2}
\left[\frac{M}{r^2}\right]\,\left[\theta(z)-1\right]=2\,\bar{U}^1\,\frac{d}{dS}\bar{U}^1-2\eta\,\left[\theta(z)-1\right]\,(z-1)\,
e^{-2\eta\,\left[\theta(z)-1\right](z-1)\,S}.
\end{equation}
By multiplying (\ref{c1}) by $2\bar{U}^1$, and comparing with (\ref{c2}), we obtain the expression
{\small \begin{equation}\label{d1}
\left[\bar{U}^1\right]^2 = 2\eta\,\left[\theta(z)-1\right]\,(z-1)\,e^{-2\eta\,\left[\theta(z)-1\right](z-1)\,S}\,
\left[\frac{(r-2M)^2}{4M+2\eta\,\left[\theta(z)-1\right](z-1)(r-2M)^2}
\right].
\end{equation}}
The equation (\ref{norma}) can be written as
\begin{equation}\label{d2}
\left[\bar{U}^1\right]^2 = e^{-2\eta\,\left[\theta(z)-1\right](z-1)\,S} - \left[\theta(z)-1\right]\,\frac{(r-2M)}{r}.
\end{equation}
Therefore, equating Eqs. (\ref{d1}) and (\ref{d2}), we obtain the expression for $r(s)$
\begin{equation}\label{rs}
r(S)= \frac{6^{1/3}\,f(S)^{1/3}}{3\,A(z)\left[\theta(z)-1\right]} + \frac{M 6^{1/3} \left[e^{-2A(z)\,S}\left[\theta(z)-1\right]\right]}{3\,f(S)^{1/3}}
+2M,
\end{equation}
where we have defined the auxiliary functions $A(z)=\eta\,(z-1)\,\left[\theta(z)-1\right]$ and $f(S)$
\begin{equation}
f(S)= M\,A^2(z)\left[\theta(z)-1\right]^2 \left[9M\,e^{-2A(z)\,S} + \sqrt{3}\,F(S)\right],
\end{equation}
with
{\small \begin{eqnarray}
&& F(S)=\left\{\frac{M}{A(z)\,(\theta(z)-1)}\left[27Me^{-4A(z)S}A(z)\left[\theta(z)-1\right]-2\left[e^{-6A(z)S}+1\right]-6\theta(z) e^{-4A(z)S}\left[\theta(z)-1\right]\right.\right. \nonumber \\
&& \left.\left.+
2\theta(z)^3+12\theta e^{-2A(z)S}-6\theta(z)\left[\theta(z)-1\right]-6 e^{-2A(z)S}\left(e^{-2A(z)S}+1\right)\right]\right\}^{1/2}.
\end{eqnarray}}
In this way, the rate $r(S)/(2M)$ tends to $1$ as $S\rightarrow\infty$, which means that the relativistic observer who is outside the SBH at the beginning, approaches the horizon but never reaches it. The relativistic component $\bar{U}^1(S)$ begins given after making the derivative of $r(S)$ in (\ref{rs}), with respect to the $4$-length $S=\left[g_{\alpha\beta}\,x^{\alpha}x^{\beta}\right]^{1/2}$, which is the affine parameter of our theory:
\begin{equation}\label{u1}
\bar{U}^1(S)= \frac{d}{dS} r(S),
\end{equation}
where $r(S)$ is given by Eq. (\ref{rs}). Therefore, using the results of this section, the function $\lambda(\left[r(S)\right]$ for a SBH will be given by [for $\bar{U}_1(S)=g_{11}\,\bar{U}^1(S)$]
\begin{equation}\label{lamm}
\lambda(S) =
\frac{3z\,\eta}{2} \left[\frac{\eta\,(2z+1)\,\left[\bar{\theta}(z)-1\right] + \left[\bar{U}_1(S)/r(S)\right]}{ (z-1) \left[\theta(z)-1\right]}\right].
\end{equation}
This depends on the $z$-parameter, which characterizes the relative strength of classical and quantum gravitational couplings in the covariant  derivative \cite{p2}. Recall that large z-values correspond to the dominance of classical gravitational fluctuations, while small values $ 0<  z\lesssim 8/9$, describe quantum fluctuations of spacetime.
\begin{figure}
    \centering
    \includegraphics[scale=0.5]{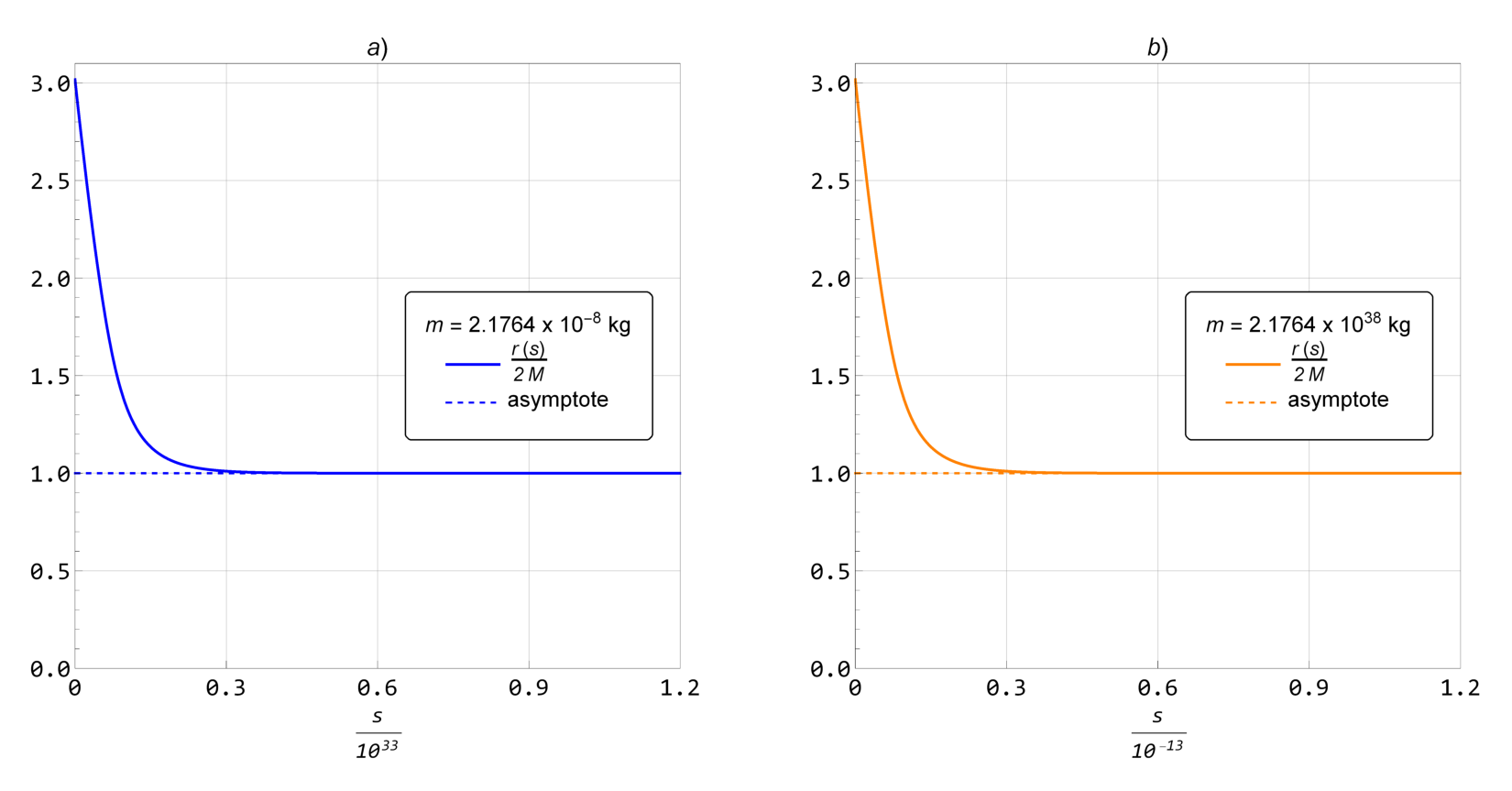}
    \caption{Evolution of $r(S)/(2M)$ from an initial value $S_0=0$ and $\eta=1/(2M)$, for a) $m=2.1764\times 10^{-8}\, {\rm kg}$, b) $m=2.1764\times 10^{38}\, {\rm kg}$.}
    \label{f2agi147}
\end{figure}

\begin{figure}
    \centering
    \includegraphics[scale=0.5]{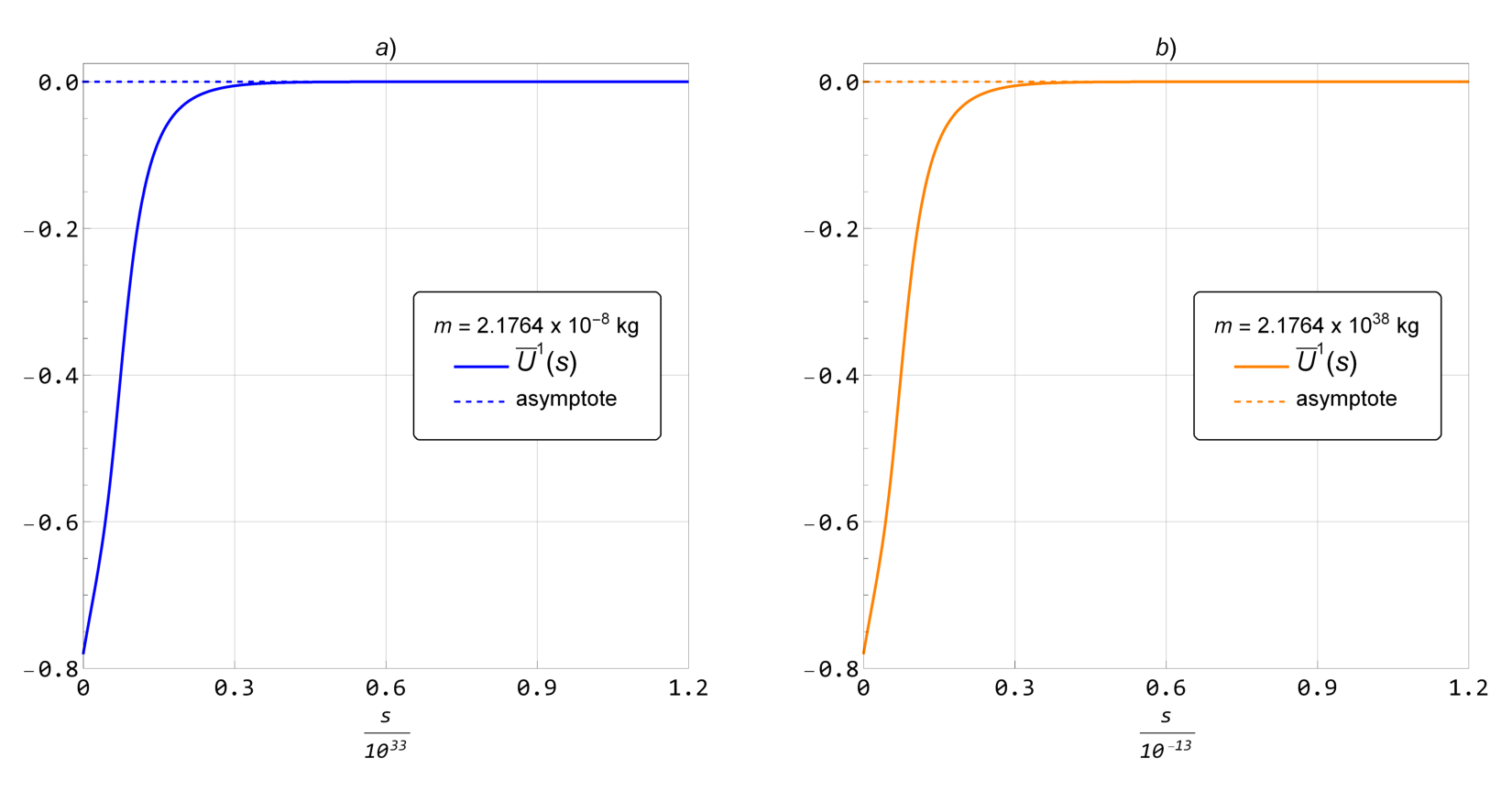}
    \caption{Evolution of $\bar{U}^1(S)\equiv \frac{dr}{dS}$, with $S_0=0$ and $\eta=1/(2M)$, for $z=(1+\sqrt{3})/2$ and a) $m=2.1764\times 10^{-8}\, {\rm kg}$, b) $m=2.1764\times 10^{38}\, {\rm kg}$.}
    \label{f3agi147}
\end{figure}

In the Figs. (\ref{f2agi147}) and (\ref{f3agi147}), we have plotted the evolution of $r(S)/(2M)$ and $\bar{U}^1(S)\equiv \frac{dr}{dS}$ from an initial value $S_0=0$, for two different Schwarzschild black holes (SBHs), with different masses and $z=(1+\sqrt{3})/2$. In all cases, we have used $\eta=1/(2M)$. The cases (\ref{f2agi147}a) and (\ref{f3agi147}a) correspond to a SBH with Planck mass $m=2.1764\times 10^{-8}\,{\rm kg}$. The cases (\ref{f2agi147}b) and (\ref{f3agi147}b) correspond to a supermassive SBH with mass $m=2.1764\times 10^{38}\,{\rm kg}$. Notice that in both cases the relativistic observer is at an initial distance: $r(S_0)/(2M) \simeq 3$, from the SBH and later approaches the Schwarzschild horizon but never reaches it. Both observers describe the same trajectory, but with characteristic times that differ by an order of $10^{46}$; the fastest being the relativistic observer who approaches the less massive BH.

\subsection{Effective gravitational potential and mass density with geometric quantum gravitational corrections included}

The effective gravitational potential outside a SBH, when the corrections associated with quantum gravity are taken into account, will depend on how the relativistic observer is moving. If we consider that the observer approaches the SBH according to the solution of the geodesic equations (\ref{u0}) and (\ref{u1}), we have that
\begin{equation}\label{v}
V(S)=-\frac{M}{r(S)} - \frac{\lambda(S)}{\kappa},
\end{equation}
with $r(S)$ given by (\ref{rs}). The second term in (\ref{v}) takes into account the geometric quantum corrections due to the flow of the gravitational perturbations through the horizon at $r_{h}=2M$.

\begin{figure}
    \centering
    \includegraphics[scale=0.5]{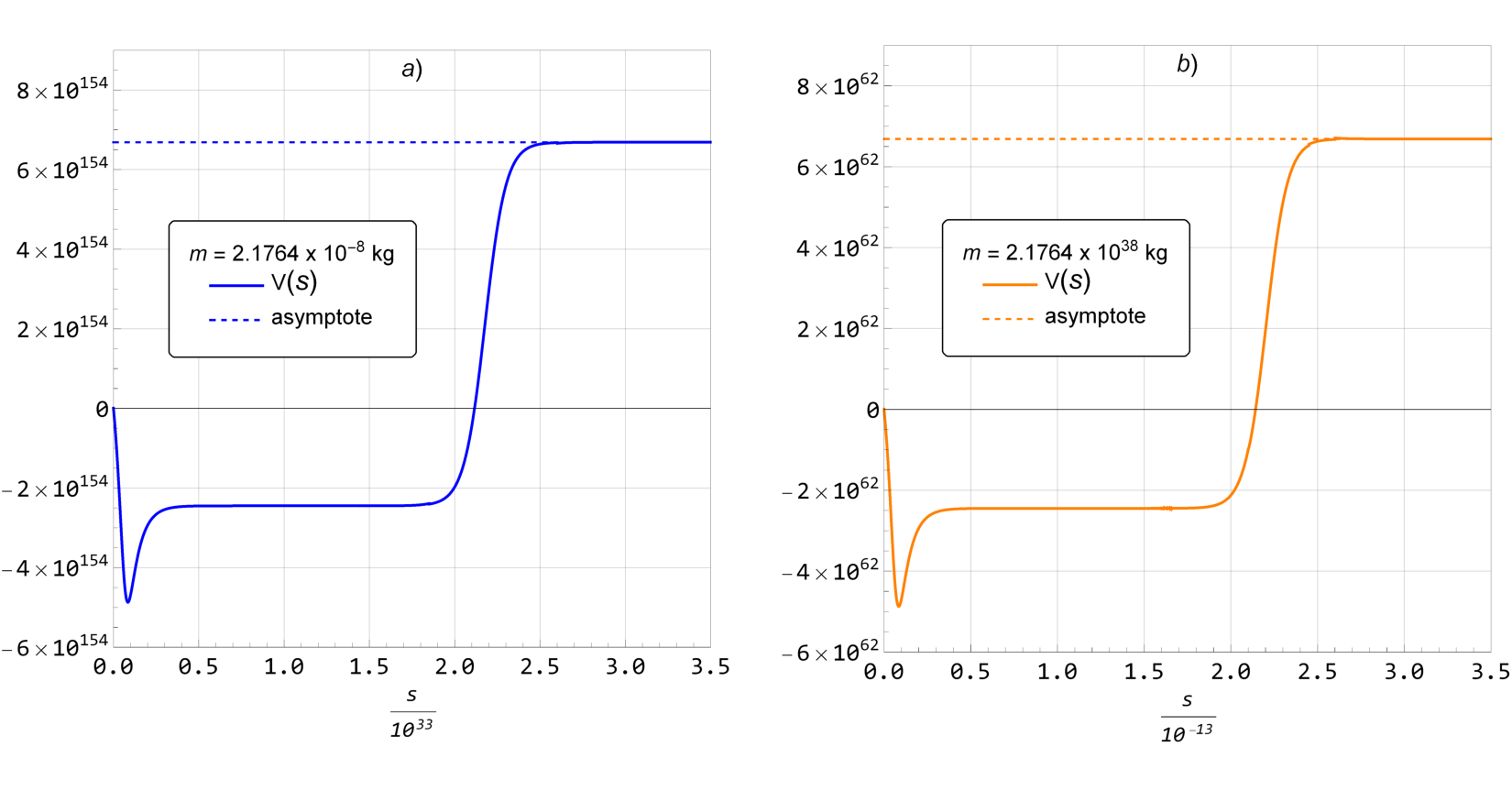}
    \caption{Effective potential $V(S)=-\frac{M}{r(S)} - \frac{\lambda(S)}{\kappa}$, with $S_0=0$ and $\eta=1/(2M)$, for $z=(1+\sqrt{3})/2$, and a) $m=2.1764\times 10^{-8}\, {\rm kg}$, b) $m=2.1764\times 10^{38}\, {\rm kg}$.}
    \label{f4agi147}
\end{figure}

In the Figs. (\ref{f4agi147}) we have plotted the effective potential $V\left[r(S)\right]=-\frac{M}{r(S)} - \frac{\lambda(S)}{\kappa}$, su\-ffered by the relativistic observer which moves with the velocities $\bar{U}^0(S)$ and $\bar{U}^1(S)$, given respectively by the expressions (\ref{u0}) and (\ref{u1}), for $z=(1+\sqrt{3})/2$. In the Fig. (\ref{f4agi147}a) we have considered the case of a Planck SBH with mass $m=2.1764\times 10^{-8}\, {\rm kg}$, and in the Fig. (\ref{f4agi147}b) we have considered a supermassive SBH with mass $m=2.1764\times 10^{38}\,{\rm kg}$. Note that, in both cases, the observer initially falls to a bound state, but then begins to climb the effective potential hill with a barrier of finite height at positive energies. This is due to the flow of spacetime fluctuations, described by the $\hat{\Omega}^{\alpha}$ field, changes its sign (from negative to positive), during the evolution of the system.

\begin{figure}
    \centering
    \includegraphics[scale=0.5]{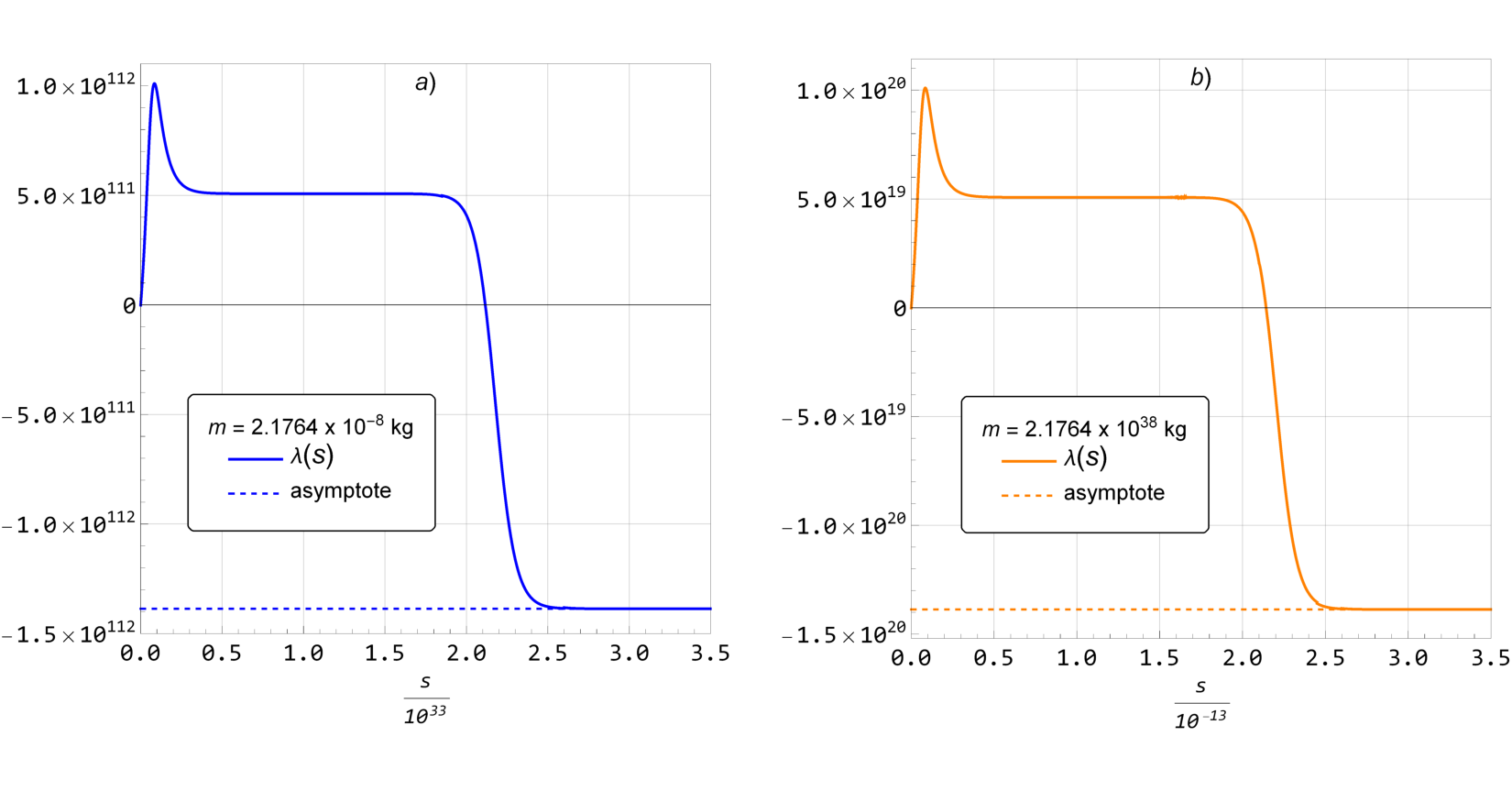}
    \caption{Evolution of $\lambda(S)$, with $S_0=0$ and $\eta=1/(2M)$, for $z=(1+\sqrt{3})/2$ and a) $m=2.1764\times 10^{-8}\, {\rm kg}$, b) $m=2.1764\times 10^{38}\, {\rm kg}$.}
    \label{f5agi147}
\end{figure}

In both cases, the relativistic velocities $\bar{U}^1(S)$ tend asymptotically to zero as the observers approaches to the horizons [see Figs. (\ref{f3agi147})], indicating that the last term in the numerator of (\ref{lamm}) cancels near the horizon. The canceled
term has a purely classical origin. However, the first term in the numerator of (\ref{lamm}) is of quantum nature and it becomes dominant near the Schwarzschild horizon. In Figs. (\ref{f5agi147}) we have plotted the evolution of the parameter $\lambda(S)$ for two SBHs: one with the Planck mass and another with a mass $10^{46}$ times the Planck mass. In both cases $\lambda(S)$ has the same behavior, being positive when the observer is far from the horizon and changing to negative when the observer approaches the Schwarzschild horizon. These values correspond respectively to negative and positive $\hat{\Omega}^{\alpha}$ fluxes. At this point the relativistic observer only detects the gravitational radiation emitted by the SBH, because it is very close to the horizon. In this situation the classical effects due to the curvature produced by the SBH are extinguished and the gravitational radiation emitted by the SBH is dominant with respect to the classical fluctuations of spacetime.

The mass density $\rho(s)$, of the collapsing system that eventually turns into SBH, will be given by the mass $m$ divided by its volume, $V_{ol}(S)=(4/3)\pi\,r^3(S)$:
\begin{equation}
\rho(S)= \frac{3\,m}{4\pi\,r^3(S)}.
\end{equation}
Both black holes increase their mass density during the evolution of the system reaching maximum asymptotic values of $1.54\times 10^{95}\,{\rm kg/m^3}$, for the SBH with Planck mass, and of $1000\,{\rm kg/m^3}$, for the SBH with a mass of $10^{46}$ times the Planck mass.

\subsection{Spectral dimensions and metric tensor with geometric quantum gravitational corrections included}

To know the spectral dimension of space-time we must calculate the elements of the metric tensor that include the quantum corrections due to space-time fluctuations caused by the radiation emitted by the SBH. The variation of the metric tensor components due to the geometric gravitational perturbations, for $\bar{U}^2=\bar{U}^3=0$ and $b=z\eta$, is given by [see Eq. (\ref{delg})]:
\begin{footnotesize}\begin{eqnarray}
&&{\delta g}_{\alpha\beta}=\left(\frac{1}{z\eta}\right)\,\left<B\right|\hat{\delta g}_{\alpha\beta}\left|B\right> =\nonumber \\
&& \left( \begin{array}{cccc}  2\left[\frac{\left[\theta(z)-1\right]}{z}\,\, g_{00}  -\,\bar{U}_{0}\bar{U}_{0}\right]  &
-2\,\bar{U}_{0}\bar{U}_{1} & 0
 & 0  \\
-2\,\bar{U}_{1}\bar{U}_{0}  &   2\left[\frac{\left[\theta(z)-1\right]}{z}\,\, g_{11}  -\,\bar{U}_{1}\bar{U}_{1}\right]   &  0 & 0  \\
0 & 0 & 2\left[\frac{\left[\theta(z)-1\right]}{z}\,\, g_{22}\right] & 0\\
0 & 0 & 0 & 2\left[\frac{\left[\theta(z)-1\right]}{z}\,\, g_{33}\right]
\end{array} \right), \label{correc}
\end{eqnarray}\end{footnotesize}
where $\bar{U}_{0}(S)=g_{00}\,\bar{U}^{0}(S)$ is given by Eq. (\ref{u0}) and $\bar{U}_{1}(S)=-\left(\frac{r(S)-2M}{r(S)}\right)\,\bar{U}^1(S)$.  The component
$\bar{U}^1(S)\equiv \frac{dr(S)}{dS}$ was plotted in Figs. (\ref{f2agi147}a) and (\ref{f2agi147}b), for SBHs with masses $m=2.1764\times 10^{-8}\, {\rm kg}$ and $m=2.1764\times 10^{38}\, {\rm kg}$, with $\eta=1/\left(2M\right)$ and $z=(1+\sqrt{3})/2$. It can be shown that for an SBH with arbitrary mass, the velocity of the relativistic observer $\bar{U}^1(S)\rightarrow{0}$, when $S \rightarrow \infty$. In that limit, $r(S) \rightarrow 2M$, when $S\rightarrow \infty$, as can be seen in Figs. (\ref{f2agi147}a) and (\ref{f2agi147}b), such that $\bar{U}_1(S) \rightarrow 0$, when $S\rightarrow \infty$. The expectation value of the total covariant metric tensor: $\bar{g}_{\alpha\beta}= g_{\alpha\beta}+{\delta g}_{\alpha\beta}$, adopts the general form [see Eq. (\ref{gbar})]:
\begin{equation}\label{covg}
\bar{g}_{\alpha\beta}=g_{\alpha\beta}\left[1+\frac{2\left[\theta(z)-1\right]}{z} \right] -2\,\bar{U}_{\alpha}\bar{U}_{\beta}.
\end{equation}
Alternatively, we can calculate the contravariant components of the total metric tensor:
\begin{equation}\label{cong}
\bar{g}^{\alpha\beta}=g^{\alpha\beta}\left[1-\frac{2\left[\theta(z)-1\right]}{z} \right] +2\,\bar{U}^{\alpha}\bar{U}^{\beta},
\end{equation}
such that the effective number of dimensions (or spectral dimensions) of spacetime is: $N(z)=\bar{g}_{\alpha\beta}\,\bar{g}^{\alpha\beta}=4\left[\theta(z)-1\right]$ \cite{p1}, for $\theta(z)$ given by Eq. (\ref{oro}).

\section{Final comments}

The introduction of classical and quantum fluctuations of spacetime relative to a Riemannian background manifold leads to alterations in the geodesic and Einstein equations. As a result, relativistic observers will describe trajectories that deviate from those in a geometry without such fluctuations [see Eq. (\ref{geo2})]. In the context of spacetime fluctuations around a SBH, the emission of Hawking radiation implies that these fluctuations are inherently quantum and that they alters the dimensionality of the background spacetime. These perturbations are also reflected in the function $\lambda(x^{\mu})$, which we obtained in equation (\ref{lamm}), and which we calculate in the vicinity of a SBH. Notably, near the Schwarzschild horizon, the contributions from classical fluctuations cancel out, and only remain quantum contributions due to the radiation emitted by the SBH. Notably, we find that the effective gravitational potential in the vicinity of the horizon is fundamentally altered by Hawking radiation. Specifically, these quantum effects prevent the potential from tending to infinitely negative values as an observer approaches the Schwarzschild horizon [see Figs. (\ref{f4agi147}a) and (\ref{f4agi147}b)].

Changes in the spectral dimension at different scales could have profound implications for our understanding of fundamental physics. In a recent work \cite{p1} it was found that in equilibrium the spectral dimension of spacetime at cosmological scales is $N(z=2)=8/(1+\sqrt{2}) \simeq 2.4721$. In this work, we have studied the spectral dimension of spacetime at short scales, which are related to the surrounding of a SBH. We have found that, when the parameter takes the value $z_*=(1+\sqrt{3})/2\simeq 1.3660$, the observer approaching the Schwarzschild horizon with $\bar{U}^0(S)$ given by (\ref{u0}), $\bar{U}^1(S)$ given by (\ref{u1}) [see Figs. (\ref{f3agi147}a) and (\ref{f3agi147}b)], and $\bar{U}^2=\bar{U}^3=0$, perceives a spectral dimension $N(z=z_*)=4\left[\theta(z_*)-1\right]=2.8849$, which is different from that found at cosmological scales in a previous work \cite{p2}. Under these conditions, all studied SBHs with masses ranging from the Planck mass to $10^{46}$ times the Planck mass present the same stability configuration. The fact that this spectral dimension is the same for SBHs of very different masses suggests an universal property of these objects under those particular conditions. Therefore, the spectral dimension is a novel way to quantify the dimensionality of spacetime, sensitive to its possible structure at different scales and used as a key tool in the investigation of quantum gravity.

\section*{Acknowledgements}

The authors acknowledge CONICET, Argentina (PIP 11220200100110CO), and UNMdP (EXA1156/24) for financial support.

\section*{Data Availability Statement}

No Data associated in the manuscript.

\end{document}